\documentclass{iopjournal}

\usepackage[utf8]{inputenc}
\usepackage{amsmath,amssymb,amsthm}
\usepackage{graphicx}
\usepackage{lmodern}
\usepackage{hyperref}
\usepackage{setspace}
\usepackage{soul}
\usepackage{xcolor}
\newtheorem{theorem}{Theorem}

\theoremstyle{definition}

\begin{document}

\title{Area Monotonicity of Wormhole Throats and a Geometric Bound on Information Transfer}

\author{Fuat Berkin Altunkaynak$^{1}$\orcid{0009-0003-8452-5129}, Aslı Tuncer$^{2}$\orcid{0000-0003-0241-9932} }

\affil{$^1$Huseyin Avni Sozen Anatolian High School, Istanbul, Turkey}

\affil{$^2$Physics, Istanbul Health and Technology University, Istanbul, Turkey}

\email{fuatberkin34@gmail.com and asli.ozdemir@istun.edu.tr}


\begin{abstract}
We develop a semiclassical geometric framework to constrain information transfer through traversable wormholes. This study is motivated by the growing intersection between spacetime geometry and quantum information theory, specifically the ER=EPR conjecture and the bit-thread formulation of holographic entropy. First, we prove a geometric monotonicity result for traversable wormhole throats, demonstrating that after a traversable window is established via an averaged null energy condition (ANEC) violating deformation, any subsequent signal-carrying matter satisfying the pointwise null energy condition (NEC) causes the throat cross-sectional area to be non-increasing. Second, we utilize this monotonicity to derive a semiclassical geometric upper bound on the number of independent quantum degrees of freedom (qubits) transmissible through the wormhole. This bound $Q_{\max} \leq A_{\min}/4G_N$, is motivated via the Max-Flow Min-Cut theorem for bit threads and interpreted as a geometric proxy for the holographic capacity of a quantum teleportation channel. We further discuss a holographic tensor-network analogy based on the HaPPY code, where the discrete max-flow/min-cut theorem provides an illustrative graph-theoretic counterpart of the bottleneck structure. Our results identify the wormhole throat as a natural geometric bottleneck, providing a geometric perspective on information-transfer limits in semiclassical gravity.
\end{abstract}

\section{Introduction}

The interplay between entanglement and geometry has been a rapidly developing area of research in modern theoretical physics. One of the most concrete physical realizations of this relationship is the correspondence between the entanglement entropy of a boundary region and the area of the minimal bulk surface homologous to the boundary region, as prescribed by Ryu-Takayanagi and Hubeny-Rangamani-Takayanagi \cite{Ryu2006,Hubeny2007}. The eternal AdS-Schwarzschild black hole provides a canonical example of this correspondence and is dual to the thermofield double state for two entangled quantum systems \cite{Maldacena2003}. This duality is one of the most natural realizations of the broader ``ER=EPR'' idea \cite{Maldacena2013} relating quantum entanglement to spacetime connectedness.

Classically, these wormholes are non-traversable; no signal entering one exterior can reach the other. This is because of the energy conditions imposed by general relativity. However, wormholes in general relativity can be traversable once negative energy is introduced \cite{Morris1988}. An eternal two-sided AdS-Schwarzschild black hole can be rendered traversable by introducing a coupling between the two boundaries that injects negative energy and violates the averaged null energy condition (ANEC), as shown by Gao, Jafferis, and Wall (GJW) \cite{Gao2017}.

The possibility of traversability raises a natural and broadly relevant question: How much information can be sent through such a wormhole at a time? Previous analyzes have addressed this question in limited cases, often relying on nearly two-dimensional geometries or perturbative backreaction arguments showing that the wormhole is closed after some information is thrown in \cite{Hirano2019,Freivogel2020,Maldacena2017}. 
However, no general area theorem for traversable wormhole throats or universal bound on their communication capacity has been rigorously established.

In this paper, we provide a concrete realization of such a result and bound. We prove that after the GJW deformation is switched off, the area of the wormhole throat, a smooth codimension-2 cross-section, cannot be increased by any infalling matter satisfying the null energy condition (NEC). This geometric monotonicity is formulated through a two-stage separation between the topology-establishing background, namely the GJW deformation that violates the ANEC and opens the traversable window, and the subsequent signal-carrying matter, which is assumed to satisfy the NEC. This separation resolves the apparent conflict of applying the NEC-based theorems to a fundamentally ANEC-violating geometry. We impose the local bottleneck condition ($\theta_{(k)}|_{S_0} \leq 0$) that characterizes the throat as a local geometric constraint, distinct from a global extremal-surface condition.

It should be noted that the geometric statement itself—that area is non-increasing under NEC-satisfying matter—follows from the standard Raychaudhuri equation. The novelty of our work lies not in the mathematical theorem per se but in its application to the physically nontrivial setting of traversable wormholes, where the two-stage separation provides the necessary conceptual bridge to apply classical focusing results to a quantum-stabilized geometry.

By combining this geometric monotonicity with the max-flow / min-cut theorem in the bit-thread formulation of holographic entropy \cite{Freedman2017}, we derive a semiclassical upper bound for information transmission: the maximum number of independent quantum degrees of freedom that can be transmitted during the traversable window is at most $A_{\min}/4G_N$. We further illustrate this capacity bound in a holographic tensor-network model built from two entangled black holes in two glued HaPPY codes, applying the max-flow/min-cut theorem in a discrete setting where the corresponding max-flow/min-cut relation becomes graph theoretically exact and provides an illustrative discrete analog of the bottleneck structure. Our results identify a throat area monotonicity statement for traversable wormhole cross-sections and use it to motivate a geometric bound on information transfer.

Although our analysis involves null congruences and an area-entropy relation, it is conceptually distinct from the Bousso covariant entropy bound \cite{Bousso2002}, which constrains entropy on light-sheets rather than the geometric capacity of a wormhole throat. Unlike Hawking's area theorem \cite{Hawking1971}, which applies to event horizons, our result constrains arbitrary wormhole throat cross-sections that act as geometric bottlenecks for information transfer.

\section{Geometric Setup and the NEC Conflict}

A major conceptual hurdle in proving area theorems for traversable wormholes is the requirement for energy condition violation. Since the throat must be supported by negative energy (ANEC violation), any theorem assuming the NEC might seem inapplicable. We resolve this potential conflict by adopting a two-pulse framework that distinguishes between the topology-establishing background and the signal-carrying matter. The background pulse, represented by $T_{ab}^{\text{bkg}}$, corresponds to the non-local GJW deformation that establishes the traversable window. Once the wormhole is rendered open, any subsequent classical or quantum matter sent through it is characterized by the stress-energy tensor $T_{ab}^{\text{sig}}$. For any realistic signal, this tensor is expected to satisfy the pointwise NEC. Our result specifically describes the geometric response of the pre-existing traversable throat to this NEC-satisfying signal. This separation is physically justified because if the signal itself violates the NEC, it would further expand the throat, potentially leading to instabilities. By assuming the NEC for the signal, we establish a conservative and robust upper bound on the channel capacity based on the pre-existing geometry.

\subsection{Definition of the Wormhole Throat}

\begin{figure}[htbp]
\centering
\includegraphics[width=0.60\textwidth]{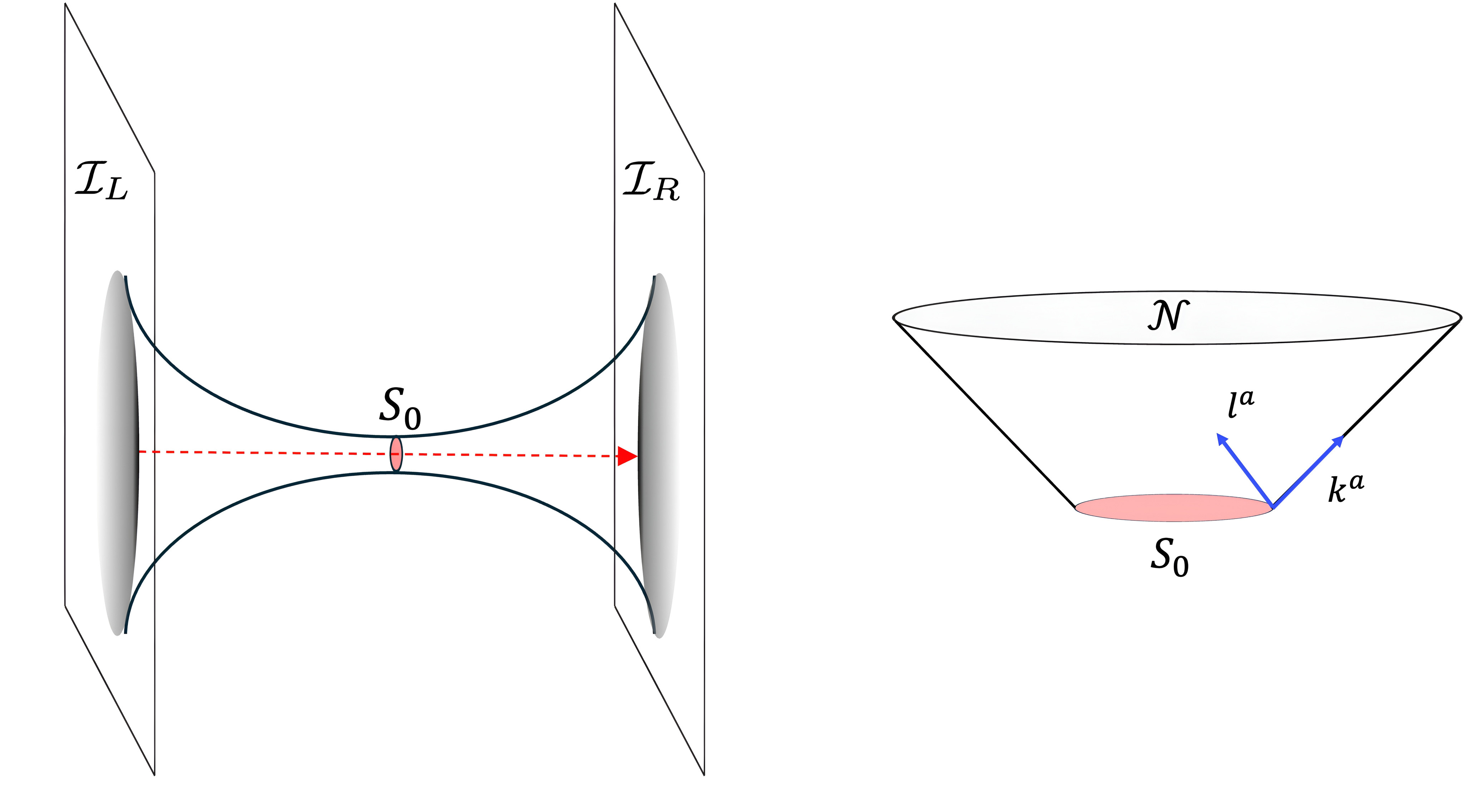}
\caption{(a) A spacelike hypersurface intersecting a traversable wormhole. The minimal cross-section $S_0$ acts as a geometric bottleneck connecting the two asymptotic regions $\mathcal{I}_L$ and $\mathcal{I}_R$. The red arrow indicates the traversable window. (b) The null hypersurface $\mathcal{N}$ generated from $S_0$ by future-directed null geodesics with tangent vectors $\ell^a$ and $k^a$.}
\label{fig:setup}
\end{figure}

We work in a smooth, time-oriented spacetime $(M, g_{ab})$ of dimension $D \geq 3$. Let $\Sigma$ be a spacelike hypersurface that intersects the wormhole region and connects two asymptotic boundaries, $\mathcal{L}$ and $\mathcal{R}$. A codimension-2 surface $S_0 \subset \Sigma$ is defined as a cross-section of the wormhole throat if it is a minimal-area surface separating the two boundaries. We define two future-directed null normals, $k^a$ (pointing in the direction of traversal) and $\ell^a$ (pointing inward), such that they satisfy the following normalization:
\begin{equation}
k^a \ell_a = -1.
\end{equation}
The induced metric on $S_0$ is given by:
\begin{equation}
h_{ab} = g_{ab} + k_a \ell_b + \ell_a k_b.
\end{equation}
The null expansion along $k^a$, denoted by $\theta_{(k)} = h^{ab} \nabla_a k_b$, characterizes the rate of change of the area element along the null congruence. We identify the throat as a geometric bottleneck if it satisfies the following condition:
\begin{equation}
\theta_{(k)}|_{S_0} \leq 0.
\end{equation}
Physically, Eq.~(3) implies that, as a signal passes through $S_0$, the surrounding null rays are either parallel or converging. This condition is weaker than that of an extremal surface, where the expansion vanishes individually, allowing us to describe dynamic, non-stationary throats. The geometric setup is illustrated in Fig.~\ref{fig:setup}.

\section{Theorem 1: Area Monotonicity in Traversable Throats}

We now formulate the precise geometric result that governs the evolution of the throat area under signal propagation.

\begin{theorem}[Geometric Monotonicity]
Let $S_0$ be a throat cross-section in a traversable AdS wormhole. Let $\mathcal{N}$ be the null hypersurface generated from $S_0$ by future-directed null geodesics with tangent $k^a$ along the traversal direction. If the signal-carrying matter satisfies the NEC ($T_{ab} k^a k^b \ge 0$) along $\mathcal{N}$, then the area $A(\lambda)$ of any cross-section $S(\lambda) \subset \mathcal{N}$ is non-increasing toward the future:
\begin{equation}
\frac{dA}{d\lambda} \leq 0.
\label{eq:areamonotonicity}
\end{equation}
\end{theorem}

\subsection{Proof via the Raychaudhuri Equation}

For a hypersurface-orthogonal null congruence, the vorticity vanishes, and the evolution of the expansion $\theta$ is governed by the Raychaudhuri equation:
\begin{equation}
\frac{d\theta}{d\lambda} = -\frac{1}{D-2}\theta^2 - \sigma_{ab}\sigma^{ab} - 8\pi G T_{ab}k^a k^b.
\end{equation}
In Eq.~(5), the term $\sigma_{ab}$ represents the shear tensor. Since the shear term $\sigma_{ab}\sigma^{ab}$ is non-negative and we assume the NEC for the signal matter ($T_{ab}k^a k^b \ge 0$), we obtain the fundamental inequality:
\begin{equation}
\frac{d\theta}{d\lambda} \leq -\frac{1}{D-2}\theta^2.
\end{equation}
This inequality implies that if the expansion starts as non-positive at the throat, as defined in Eq.~(3), it must remain non-positive for all future $\lambda > 0$ until a conjugate point is reached. Given that the area element evolves according to $\frac{d\sqrt{h}}{d\lambda} = \theta \sqrt{h}$, the total area $A(\lambda)$ satisfies the integral relation:
\begin{equation}
\frac{dA}{d\lambda} = \int_{S(\lambda)} \theta(\lambda, x) \sqrt{h(\lambda, x)} \, d^{D-2}x.
\end{equation}
Because the integrand in Eq.~(7) is non-positive everywhere on the surface, the derivative of the area is necessarily non-positive, thus proving Eq.~(4). The equality holds only if the congruence is shear-free and there is no null energy flux along the generators.\qed

\subsection{Physical Significance: Throat vs. Horizon}
\label{sec:TvsH}

It is crucial to distinguish this result from Hawking's area theorem~\cite{Hawking1971}. While Hawking's theorem applies to event horizons under appropriate assumptions, including the presence of trapped surfaces, our result concerns traversable wormhole throats, which instead behave as geometric bottlenecks. This suggests that the initial area of the throat $A_{\min}$ represents the natural maximum aperture available for communication. Any subsequent NEC-satisfying signal does not expand this window, but propagates through a bottleneck that is geometrically stable or shrinking.

\section{Main Result: The Information Capacity Bound}

\subsection{Bit Threads and the Max-Flow/Min-Cut Picture}

To relate the geometric monotonicity result to information transfer, we utilize the bit-thread formulation of the holographic entropy \cite{Freedman2017}. In this formulation, bit threads are divergence-free vector fields $v^a$ that satisfy
\begin{equation}
\nabla_a v^a = 0, \quad |v| \leq \frac{1}{4G_N}.
\end{equation}
The max-flow min-cut theorem states that the maximum flux of such a vector field between the boundary regions is equal to the area of the minimal bulk surface separating them divided by $4G_N$:
\begin{equation}
\text{Max Flux} = \max_v \int_{\mathcal{L}} v \cdot dA = \frac{A_{\min}}{4G_N}.
\end{equation}

In the present setting, the relevant minimal surface is the wormhole throat cross-section acting as a geometric bottleneck between the two asymptotic regions. Theorem 1 implies that once the traversable window has been established and the ANEC-violating deformation has been switched off, subsequent NEC-satisfying signal matter cannot increase the cross-sectional area available for transmission. This identifies the throat as a geometric bottleneck whose initial area provides a natural upper scale for the amount of information that can pass through the wormhole during the traversable interval.

\subsection{Proposed Semiclassical Capacity Bound}
We therefore propose the following semiclassical geometric bound on information transfer through a traversable wormhole:
\begin{equation}
Q_{\max} \leq \frac{A_{\min}}{4G_N}.
\label{eq:ineqQ}
\end{equation}
Here, $Q_{\max}$ denotes the maximum number of independent quantum degrees of freedom that can be transmitted through the wormhole during the traversable window, while $A_{\min}$ is the minimal throat area identified in Sec.~\ref{sec:TvsH}, which now plays the role of a geometric bottleneck setting the upper scale for information transfer.

In the holographic dictionary, information transfer through a traversable wormhole may be interpreted in terms of a quantum teleportation protocol. Teleporting quantum information between the two boundaries requires prior shared entanglement~\cite{Bennett1993}, which is geometrically represented by bit threads in the bulk~\cite{Freedman2017}. Related information-theoretic quantities, such as conditional mutual information 
$I(\mathcal{L}:\mathcal{R}\mid M)$ between the two boundaries conditioned on the infalling sector $M$, may provide suggestive diagnostics of traversable information transfer, although we do not pursue this direction here.

This inequality should be understood as a semiclassical geometric bound rather than as a fully rigorous quantum channel-capacity theorem. In the bit-thread picture, the maximal information flux between the two asymptotic boundaries is controlled by the minimal separating surface in the bulk. The wormhole throat therefore acts as a geometric bottleneck that limits the amount of transmissible boundary information during the traversable interval.

The distinction between the background deformation that violates the ANEC and the subsequent signal matter that satisfies the NEC is essential in this interpretation. Once the traversable window has been established, Theorem~1 implies that the signal cannot enlarge the available throat cross-section. Consequently, the information-transfer scale is bounded by the pre-existing bottleneck geometry rather than dynamically enhanced by the signal itself.

This bound is conceptually distinct from both Hawking's area theorem and the Bousso covariant entropy bound. Rather than constraining event horizons or light-sheets, Eq.~(\ref{eq:ineqQ}) characterizes traversable wormhole throats as local geometric constraints on communication between two asymptotic regions.

In this sense, the proposed bound is loosely reminiscent of area-based entropy bounds in gravitational systems, although its interpretation here is operational rather than thermodynamic. 

\section{Discussion}

The main result of this work is the semiclassical monotonicity statement for traversable wormhole throats and its interpretation as a geometric constraint on information transfer. The proof relies only on the local focusing behavior of null congruences under NEC-satisfying signal matter. The nontrivial point is therefore not the Raychaudhuri equation itself, but its application to a traversable wormhole geometry in which the ANEC-violating deformation that opens the wormhole is separated from the later signal-carrying matter.

This separation also clarifies the role of the energy conditions. A traversable wormhole requires an ANEC-violating ingredient in order to become open, but this does not imply that all subsequent matter sent through the wormhole must violate the NEC. Once the traversable window exists, ordinary signal matter satisfying the pointwise NEC can propagate through it. Our theorem shows that such matter cannot enlarge the throat area; instead, it propagates through a bottleneck that is geometrically stable or shrinking. In this sense, the throat area acts as a natural semiclassical scale controlling the amount of information that can be transmitted during the traversable interval.

The proposed bound should therefore be interpreted conservatively. It is not claimed to provide a complete proof of quantum channel capacity in a dynamical gravitational spacetime. Rather, Eq.~(\ref{eq:ineqQ}) identifies the wormhole throat as a geometric structure that restricts communication between the two asymptotic regions. A fully rigorous channel-capacity theorem would require a more detailed treatment of encoding, decoding, noise, backreaction, and the operational structure of the boundary quantum channel.

\emph{Discrete tensor-network analogy.}

The same bottleneck logic also admits a natural discrete interpretation in holographic tensor-network models such as the HaPPY code~\cite{Pastawski2015}. In such models, the entanglement between two boundary regions is controlled by the minimal cut $(\gamma_{min})$ that separates the corresponding sectors of the tensor-network. The graph-theoretic max-flow / min-cut theorem then implies that the maximum number of independent units that can be routed between the two boundaries is bounded by the number $(n_{min})$ of bonds crossing this cut.

In this sense, the minimal cut $(\gamma_{min})$ plays the role of a discrete bottleneck for information flow, analogous to the throat cross-section $(S_0)$ in the continuum geometry. The comparison is particularly suggestive because both settings relate information transfer to a minimal separating structure: in the continuum description, this structure is a codimension-2 throat surface, while in the tensor-network setting it is a graph-theoretic minimal cut. The tensor-network picture therefore provides an intuitive realization of the same geometric bottleneck principle underlying Eq.~(\ref{eq:ineqQ}). In the HaPPY code, each bond crossing the minimal cut carries one unit of entanglement entropy, so $n_{\min}$ represents the discrete counterpart of the continuum capacity scale $A_{\min}/4G_N$ in Eq.~(\ref{eq:ineqQ}). This identification shows that the tensor-network model provides a concrete realization in which the capacity bound becomes exactly $Q_{\max}=n_{\min}$, supporting the geometric interpretation of our semiclassical result.

At the same time, this analogy should not be interpreted too strongly. The HaPPY code does not reproduce the Raychaudhuri equation, the NEC, or the dynamical geometry of a traversable wormhole. Its role is more limited and primarily interpretive: it illustrates how area-like bottlenecks naturally constrain information flow where the max-flow/min-cut relation becomes exact. In this sense, the tensor-network construction may be viewed as a useful consistency check and an intuitive discrete counterpart of the continuum argument, rather than as an independent derivation of the gravitational bound.

\emph{Relation to information recovery.}

The proposed bound has a particularly natural interpretation in the context of black-hole information recovery and Hayden--Preskill-type protocols. At the Page time, the remaining black hole and the emitted radiation have comparable entropy, so the number of quantum degrees of freedom required, in principle, to reconstruct the interior is set by the same area scale that controls the holographic entropy of the black hole. In this regime, a traversable wormhole provides a geometric channel through which interior information may be transferred between the two entangled boundary systems.

From this perspective, Eq.~(10) suggests a direct geometric constraint on Page-time recovery. If the traversable channel saturates the throat-area bound, then the maximum amount of quantum information that can pass through the wormhole is of order
\[
Q_{\max} \sim \frac{A_{\min}}{4G_N},
\]
which is precisely the entropy scale associated with the black-hole degrees of freedom accessible at Page time, up to the conventional factor of \(\ln 2\) relating the entropy to a qubit count. Thus, the throat area does not merely bound an abstract communication channel; it bounds the amount of information that would be needed, in principle, to recover the interior degrees of freedom compatible with unitary black-hole evolution.

This provides a geometric realization of the idea that information recovery is possible without fundamental information loss: the traversable wormhole channel can, in principle, transmit the Page-time amount of quantum information, while the throat area determines the maximal scale of that transmission. However, the result should be interpreted as a semiclassical capacity statement rather than as a complete reconstruction theorem. We do not prove a full decoding protocol for the black-hole interior. Instead, we identify the wormhole throat as the geometric bottleneck whose area controls whether enough quantum information can pass through the traversable channel to support unitary recovery in principle.

\section{Conclusion}

We have shown that traversable wormhole throats obey a natural geometric monotonicity property under NEC-satisfying signal propagation. By separating the ANEC-violating deformation responsible for opening the traversable window from the subsequent signal-carrying matter, we demonstrated that the cross-sectional area of the throat cannot increase along the relevant null congruence.

This monotonicity identifies the wormhole throat as a geometric bottleneck for information transfer. Motivated by the bit-thread formulation of holographic entropy, we proposed the semiclassical bound
\begin{equation}
Q_{\max} \leq \frac{A_{\min}}{4G_N},
\end{equation}
which relates the amount of transmissible quantum information to the minimal throat area available during the traversable interval. The bound should be interpreted as a geometric constraint on communication in semiclassical gravity rather than as a fully rigorous quantum channel-capacity theorem.

More broadly, our results suggest that traversable wormhole geometries provide a natural setting in which geometric focusing properties and information-theoretic constraints become closely connected. In this picture, the wormhole throat emerges as a local geometric structure that limits the flow of quantum information between asymptotic regions.

Future work may extend these ideas to quantum-corrected settings that involve generalized entropy.
\begin{equation}
S_{\rm gen}
=
\frac{\mathrm{Area}(\gamma)}{4G_N}
+
S_{\rm bulk}(\gamma),
\end{equation}
where bulk entanglement contributions modify the effective bottleneck geometry. Establishing a corresponding monotonicity principle for quantum extremal or quantum-corrected throat surfaces may provide a natural connection with the quantum focusing conjecture and with more complete formulations of information transfer in semiclassical gravity.

\ack{We thank the anonymous referees for their valuable comments and suggestions.}

\funding{No external funding was received for this work.}

\roles{Fuat Berkin Altunkaynak: Formal analysis (lead), Investigation (lead), Writing -- original draft (supporting). Aslı Tuncer: Conceptualization (lead), Methodology (lead), Writing -- review \& editing (lead), Visualization (lead), Supervision (lead).}

\data{No new data were created or analyzed in this study. All results follow from established geometric theorems and publicly available literature.}

\suppdata{No supplementary material is provided.}

\bibliographystyle{unsrt}
\bibliography{Whref}

@article{Ryu2006,
  author = {Ryu, Shinsei and Takayanagi, Tadashi},
  title = {Holographic Derivation of Entanglement Entropy from the anti–de Sitter Space/Conformal Field Theory Correspondence},
  journal = {Phys. Rev. Lett.},
  volume = {96},
  pages = {181602},
  year = {2006},
  doi = {10.1103/PhysRevLett.96.181602}
}

@article{Hubeny2007,
  author = {Hubeny, Veronika E. and Rangamani, Mukund and Takayanagi, Tadashi},
  title = {A covariant holographic entanglement entropy proposal},
  journal = {JHEP},
  volume = {07},
  pages = {062},
  year = {2007},
  doi = {10.1088/1126-6708/2007/07/062}
}

@article{Maldacena2003,
  author = {Maldacena, Juan M.},
  title = {Eternal black holes in anti-de Sitter},
  journal = {JHEP},
  volume = {04},
  pages = {021},
  year = {2003},
  doi = {10.1088/1126-6708/2003/04/021}
}

@article{Maldacena2013,
  author = {Maldacena, Juan and Susskind, Leonard},
  title = {Cool horizons for entangled black holes},
  journal = {Fortsch. Phys.},
  volume = {61},
  pages = {781},
  year = {2013},
  doi = {10.1002/prop.201300020}
}

@article{Morris1988,
  author = {Morris, Michael S. and Thorne, Kip S.},
  title = {Wormholes in spacetime and their use for interstellar travel: A tool for teaching general relativity},
  journal = {Am. J. Phys.},
  volume = {56},
  pages = {395},
  year = {1988},
  doi = {10.1119/1.15620}
}

@article{Gao2017,
  author = {Gao, Ping and Jafferis, Daniel L. and Wall, Aron C.},
  title = {Traversable wormholes via a double trace deformation},
  journal = {JHEP},
  volume = {12},
  pages = {151},
  year = {2017},
  doi = {10.1007/JHEP12(2017)151}
}

@article{Hirano2019,
  author = {Hirano, Shinji and Lei, Yuki and van Leuven, Simon},
  title = {Information transfer and black hole evaporation via traversable BTZ wormholes},
  journal = {JHEP},
  volume = {10},
  pages = {070},
  year = {2019},
  doi = {10.1007/JHEP10(2019)070}
}

@article{Freivogel2020,
  author = {Freivogel, Ben and others},
  title = {Traversable wormholes in four dimensions},
  journal = {JHEP},
  volume = {01},
  pages = {050},
  year = {2020},
  doi = {10.1007/JHEP01(2020)050}
}

@article{Maldacena2017,
  author = {Maldacena, Juan and Stanford, Douglas and Yang, Zhenbin},
  title = {Diving into traversable wormholes},
  journal = {Fortsch. Phys.},
  volume = {65},
  pages = {1700034},
  year = {2017},
  doi = {10.1002/prop.201700034}
}

@article{Freedman2017,
  author = {Freedman, Michael and Headrick, Matthew},
  title = {Bit threads and holographic entanglement},
  journal = {Commun. Math. Phys.},
  volume = {352},
  pages = {407},
  year = {2017},
  doi = {10.1007/s00220-016-2796-3}
}

@article{Bousso2002,
  author = {Bousso, Raphael},
  title = {The holographic principle},
  journal = {Rev. Mod. Phys.},
  volume = {74},
  pages = {825},
  year = {2002},
  doi = {10.1103/RevModPhys.74.825}
}

@article{Hawking1971,
  author = {Hawking, S. W.},
  title = {Gravitational Radiation from Colliding Black Holes},
  journal = {Phys. Rev. Lett.},
  volume = {26},
  pages = {1344},
  year = {1971},
  doi = {10.1103/PhysRevLett.26.1344}
}

@article{Bennett1993,
  author = {Bennett, Charles H. and Brassard, Gilles and Cr{\'e}peau, Claude and Jozsa, Richard and Peres, Asher and Wootters, William K.},
  title = {Teleporting an Unknown Quantum State via Dual Classical and Einstein-Podolsky-Rosen Channels},
  journal = {Phys. Rev. Lett.},
  volume = {70},
  pages = {1895--1899},
  year = {1993}
}

@article{Pastawski2015,
  author = {Pastawski, Fernando and Yoshida, Beni and Harlow, Daniel and Preskill, John},
  title = {Holographic quantum error-correcting codes: toy models for the bulk/boundary correspondence},
  journal = {JHEP},
  volume = {06},
  pages = {149},
  year = {2015},
  doi = {10.1007/JHEP06(2015)149}
}

\end{document}